\documentclass[sigconf, authorversion, nonacm]{acmart}

\usepackage{threeparttable}

\begin{document}

\title{Multi-Stage Field Extraction of Financial Documents \\ with OCR and Compact Vision-Language Models}

\author{Yichao Jin}
\email{jinyichao@ocbc.com}
\affiliation{%
  \institution{OCBC, Singapore}
  \city{}
  \country{}
}

\author{Yushuo Wang}
\email{yushuowang@ocbc.com}
\affiliation{%
  \institution{OCBC, Singapore}
  \city{}
  \country{}
}

\author{Qishuai Zhong}
\email{qishuaizhong@ocbc.com}
\affiliation{%
  \institution{OCBC, Singapore}
  \city{}
  \country{}
}

\author{Kent Chiu Jin-Chun}
\email{kentchiu@ocbc.com}
\affiliation{%
  \institution{OCBC, Singapore}
  \city{}
  \country{}
}

\author{Kenneth Zhu Ke}
\email{kennethzhu@ocbc.com}
\affiliation{%
  \institution{OCBC, Singapore}
  \city{}
  \country{}
}

\author{Donald MacDonald}
\email{donaldrm@ocbc.com}
\affiliation{%
  \institution{OCBC, Singapore}
  \city{}
  \country{}
}

\begin{abstract}
Financial documents are essential sources of information for regulators, auditors, and financial institutions, particularly for assessing the wealth and compliance of Small to Medium sized Businesses (SMBs). However, these documents are often difficult to parse. They are rarely born-digital and instead are distributed as scanned images that are not machine readable. The scans themselves are low in resolution, affected by skew or rotation, and often contain noisy backgrounds. They also tend to be heterogeneous, mixing narratives, tables, figures, and multilingual content even within the same report. Such characteristics pose major challenges for structured and automated information extractions.

In this paper, we propose a multistage pipeline that leverages traditional image processing models and OCR extraction, together with compact VLMs for structured field extraction of large-scale financial documents. Our approach begins with image pre-processing, including segmentation, orientation detection, and size normalization. Multilingual OCR is then applied to recover page-level text. Upon analyzing the text information, pages are retrieved for coherent sections such as financial statements, board compensation, company background. Finally, compact VLMs are operated within these narrowed-down scopes to extract structured financial indicators. In this way, we significantly reduce computational overhead while remarkably improving extraction accuracy and interpretability.

Our approach is evaluated using an internal corpus of multi-lingual, scanned financial documents (including company financial statements and 3rd-party auditing reports). The results demonstrate that compact VLMs, together with a multistage pipeline, can achieve \textbf{8.8 times} higher field-level accuracy relative to directly feeding the whole document into large VLMs, only at a tiny \textbf{0.7\%} of the GPU cost and \textbf{92.6\%} less end-to-end service latency. Our results highlight the value of modular, multi-stage pipelines for practical financial document understanding.
\end{abstract}





\maketitle

\section{Introduction}
Financial documents such as annual reports, earnings statements, and tax filings provide critical evidence of the financial status of Small and Medium-sized Businesses (SMBs). Regulators, auditors, and financial institutions depend on structured data from these reports (including revenue, expenses, and declared assets) to support compliance checks, credit evaluations, and risk assessments. However, extracting such structured information automatically and reliably remains a challenge, even with recent advances in Large Language Models (LLMs) and Vision–Language Models (VLMs).

Several factors contribute to this challenge
\begin{itemize}
    \item \textbf{Document length and scale:} Finacial reports often exceed dozens (if not hundreds) of pages, making naive end-to-end parsing directly using large VLMs infeasible.
    \item \textbf{Input quality:} Many SMB reports exist only as low-resolution scans, often with skewed pages and noisy image background. Therefore, this introduces difficulties with transcribing errors and misinterpretation of the layout.
    \item \textbf{Multilingual content:} For regional and global banking services, the received SMB reports could be published across various countries, with text appearing in English, Chinese, Malay, Tamil, and many other languages.
    \item \textbf{Structural heterogeneity:} Different sections cover narrative descriptions, governance details, and tabular financial statements, often with domain-specific formatting. 
\end{itemize}

Recent advances in large VLMs show promise for document understanding, but their high computational requirements limit scalability in real-world financial pipelines. Furthermore, long documents with mixed content would easily exceed model's context windows, leading to degraded or even zero extraction accuracy.

To address these limitations, we introduce an efficient multistage parsing framework built around compact VLMs. Rather than applying a large model directly to entire reports, our pipeline follows a sequence of pre-processing, transcribing, retrieval, and extracting.
\begin{itemize}
    \item \textbf{Image pre-processing} performs page segmentation, orientation angle detection, and size normalization to improve the quality of scanned pages.
    \item \textbf{Optical Character Recognition (OCR) transcription} recovers text from pre-processed images, handling multiple languages on text bounding box detection, text recognition with associated confidence scores.
    \item \textbf{Page Retrieval} incorporate a BM25-based keyword retrieval step to quickly identify pages likely to contain target information such as revenue, profit, dividends, board compensation, or company background. 
    \item \textbf{Compact VLM extraction} applies small vision-language models within these narrowed-down pages for each section to extract target structured financial fields such as revenue, profit, dividends, salary, etc.
\end{itemize}

This design offers three main advantages including 
\begin{itemize}
    \item \textbf{Efficiency:} Smaller VLMs are sufficient to achieve high accuracy when focused on relevant subsets of the documents.
    \item \textbf{Robustness:} preprocessing and OCR-based transcription mitigate the noise, and page retrieval process.
    \item \textbf{Interpretability:} Extracted values can be traced back to specific pages and sections, with transcription confidence scores to facilitate human-in-the-loop verifications.  
\end{itemize}

We validate our approach based on an internal multilingual data set consisting of company financial statements and third-party audit reports. The evaluation results indicate that compact VLMs, together with a multistage pipeline, can achieve \textbf{8.8 times} higher field-level accuracy comparing to directly feeding the whole documents into large VLMs, only at \textbf{0.7\%} of the GPU cost and \textbf{92.6\%} shorter service latency.  

These efficiency gains are particularly important in in-house financial processing environments dealing with highly confidential clients' data, where GPU resources are limited but workloads are substantial. Large financial institutions, such as OCBC, have to handle tens of thousands pages daily under strict regulatory or business deadlines, often with multiple services competing for the same hardware. In this setting, reducing computational cost and latency is not only desirable but essential for scalable and sustainable deployment of automated document understanding systems.

\section{Related Work} 
\label{section:related_work}

The task of extracting structured information from financial reports intersects several research areas, including OCR and image preprocessing, document layout understanding, financial NLP, and VLMs. Prior work has made progress in each of these areas, but challenges remain in handling long, noisy, and multilingual financial disclosures. We summarize the most relevant advances below.

\subsection{OCR and Preprocessing}
OCR remains the foundation for processing scanned financial documents. Classical engines such as Tesseract \cite{smith2007overview} are still widely used, but recent deep learning approaches have pushed accuracy further, especially for multilingual and low-quality scans \cite{du2020ppocr, li2023trocr}. Preprocessing techniques such as page segmentation \cite{chen2017convolutional}, skew correction \cite{akhter2020improving}, and image enhancement \cite{anvari2021survey} continue to improve the robustness of OCR in real-world settings. These methods are crucial for noisy documents, where small improvements in OCR can significantly impact downstream information extraction.

\subsection{Text and Layout Understanding}
Modern document understanding combines text, layout, and vision signals. LayoutLM \cite{xu2020layoutlm} and its successors \cite{xu2021layoutlmv2, huang2022layoutlmv3} have established strong baselines by jointly modeling text and layout. More recent work extended these ideas with multi-modal transformers such as LiLT \cite{wang2022lilt} and DocLLM \cite{li2023docllm}. These works demonstrate that structural cues are critical for parsing heterogeneous documents.

\subsection{Financial Document Analysis}
Financial NLP has a long history, ranging from textual analysis and extraction to sentiment modeling. Early work mostly focused on rule-based methods \cite{sheikh2012rule, im2015rule}, while more recent studies explore multimodal financial analysis that combines text with tables and figures \cite{chen2021finqa, singh2024finqapt}. However, most prior work assumes machine-readable inputs, limiting their applicability to scanned  reports. Relatively little research addresses the joint challenges of multilingual OCR, structural page retrieval, and financial field extraction.

\subsection{Vision-Language Models}
VLMs unify visual and textual reasoning for complex documents. Architectures such as Donut \cite{kim2022donut}, PaLI \cite{chen2022pali}, and UReader \cite{ye2023ureader} showed strong results on document parsing benchmarks. However, they are computationally expensive and struggle with long documents that exceed context windows. Recent works explore compact alternatives such as LayoutGPT \cite{feng2023layoutgpt}, mPLUG-Owl \cite{hu2024mplug, hu2025mplug}, and retrieval-augmented strategies \cite{mei2025survey}. Our work extends this line by combining lightweight VLMs with classical preprocessing and page retrieval, enabling efficient and robust extraction from long, scanned, noisy, and multilingual financial reports.

\section{Problem Statement and Methodology}
\label{section:problem_definition}
The goal of this section is to formalize the challenges of parsing SMB financial documents and motivate the design of an efficient multistage framework. In particular, we first formulate the problem statement with key metrics to seek after. Then we introduce a modular pipeline consisting of pre-processing, OCR transcription, page retrieval, and compact VLM extraction. Each component addresses a specific bottleneck in accuracy or efficiency, enabling scalable and robust structured information extraction.

\subsection{Problem Statement}
Formally, given a financial document $D = \{p_1, p_2, \dots, p_n\}$ consisting of $n$ scanned pages, the objective is to extract a structured set of target fields: $F = \{f_1, f_2, \dots, f_m\}$,
where each $f_i$ corresponds to a financial attribute such as revenue, profit, dividends, or executive compensation. Each page $p_i$ may contain a mixture of narrative text, tabular data, and figures in one or more languages. 

\begin{figure*} [t]
  \centering
  \includegraphics[width=\linewidth]{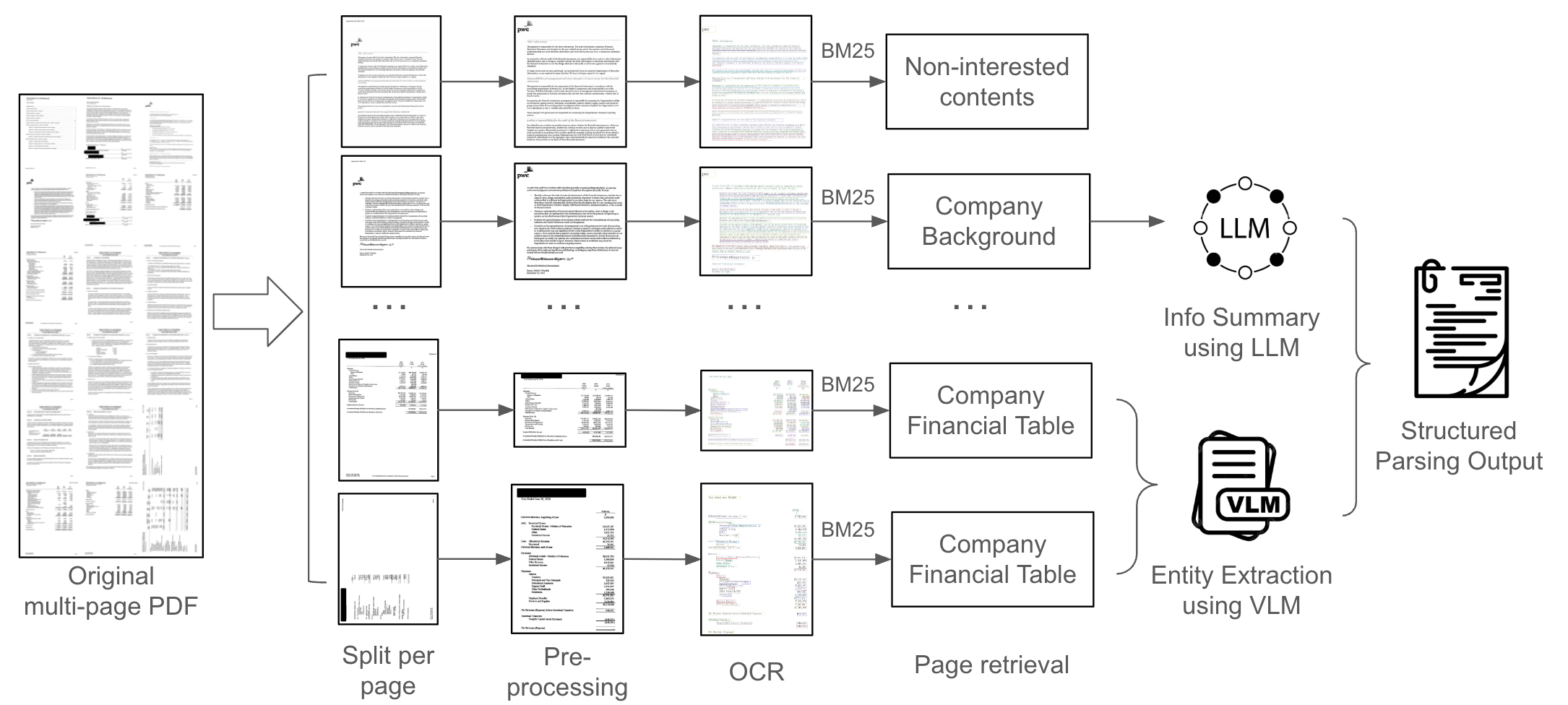}
  \caption{Overview of the proposed multi-stage parsing framework. 
    It begins with page-level splitting and image pre-processing, followed by multilingual OCR transcription. A BM25-based retrieval step filters relevant pages into semantic categories such as company background, financial tables, etc. Depending on the category, either an LLM is applied for information summarization or a compact VLM is used for pre-defined entity extraction. The outputs are merged into structured data.}
  \Description{Pipeline}
  \label{fig:pipeline}
\end{figure*}

The system must (i) recover text from noisy and potentially skewed images, (ii) identify semantically relevant subsets of pages for each $f_i$, and (iii) extract accurate field values while minimizing computational cost. 

The key performance metrics including (i) \textbf{Accuracy}: Robust extraction despite low-quality scanning, multilingual text, and heterogeneous layouts. (ii) \textbf{Efficiency}: Scalability to hundreds of pages per document under limited GPU resources. (iii) \textbf{Interpretability}: Clear traceability of extracted fields back to specific document page.

\subsection{Multi-stage Parsing Pipeline}
Figure~\ref{fig:pipeline} illustrates the proposed multi-stage parsing framework for financial documents. First, each input document is split into individual pages and passed through a series of pre-processing operations to trim non-textual areas, correct orientation, normalize size, and enhance readability. Optical Character Recognition (OCR) is then applied to transcribe multilingual text, producing text tokens with bounding box locations and confidence scores. A BM25-based retrieval module filters the pages according to their relevance to target fields, such as company background narratives or financial tables. Depending on the page type, either a Large Language Model (LLM) is used to generate text summaries (e.g., company descriptions) or a compact Vision-Language Model (VLM) is employed for fine-grained entity extraction (e.g., revenue, profit, or dividends). The results from both branches are integrated into a structured parsing output that provides interpretable financial information.

The specific steps of this pipeline, including image pre-processing, OCR transcription, page retrieval, and compact VLM extraction, are detailed in the following subsections.

\subsection{Image Pre-processing}
\begin{figure}
  \centering
  \includegraphics[width=\linewidth]{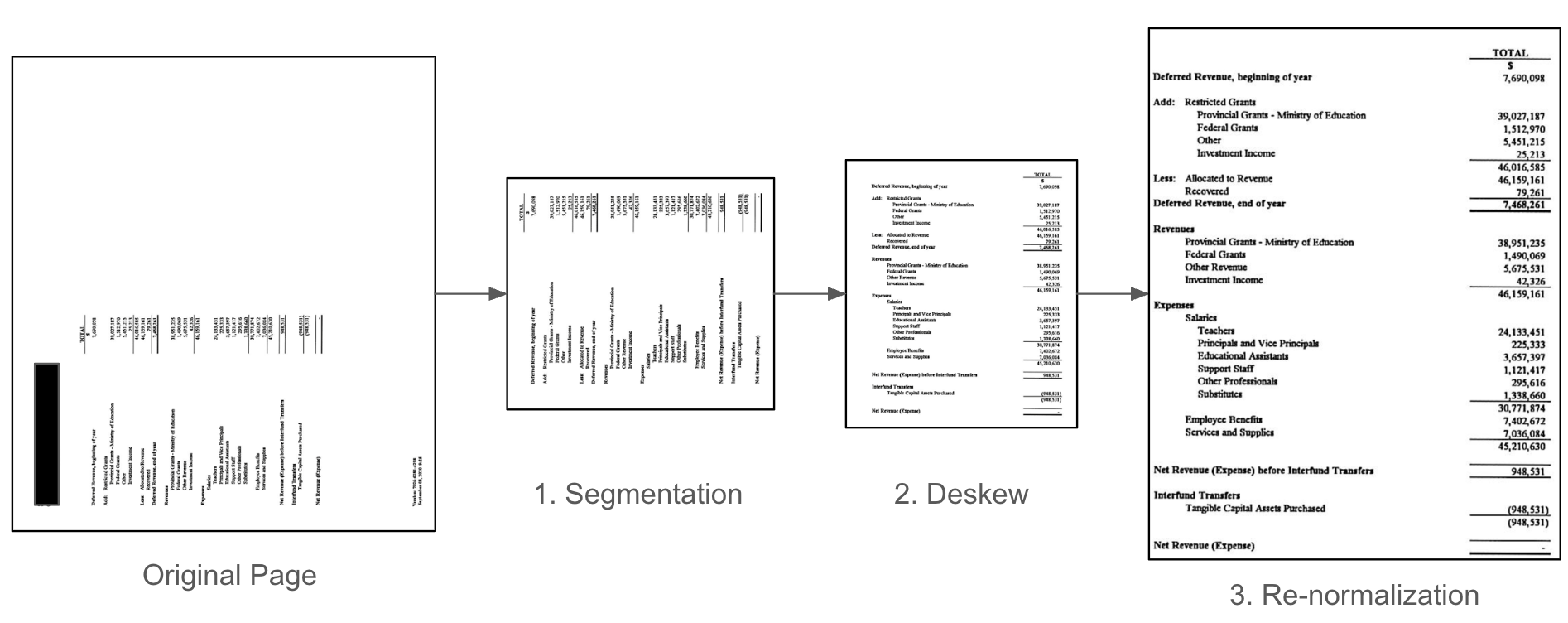}
  \caption{The illustration of image pre-processing task. It consists of three steps, including page segmentation, deskew and rotation correction, and image re-normailzation. These steps are critical to the performance of the downstream OCR and information extraction tasks.}
  \Description{Pre-processing}
  \label{fig:preprocessing}
\end{figure}

The goal of the pre-processing stage is to transform noisy scanned pages into clean, normalized inputs that maximize OCR accuracy. As illustrated in Figure \ref{fig:preprocessing}, our pipeline consists of three main steps:

\textbf{Segmentation:} Many financial reports contain large blank regions, marginal notes, or extraneous borders. We first use OpenCV’s edge detection and contour-finding algorithms \cite{xie2013image} to identify the content-bearing region of each page. This allows us to crop documents to their relevant text and tables while discarding unnecessary whitespace, which both reduces input size and magnifies small characters that would otherwise be unreadable.  

\textbf{Deskew and rotation correction:} Scanned reports are frequently misaligned, sometimes rotated by 90°, 180°, 270°, or any other random degree. We resolve this issue in two steps. First, we first apply PaddleOCR’s document orientation classifier using PP-LCNet \cite{cui2021pp} to predict the dominant page rotation into four categories (0, 90, 180, 270). After coarse alignment, we refine residual skew using the Hough transform \cite{ahmad2021efficient}, detecting the predominant angles of the text lines and rotating them to horizontal alignment. This two-stage approach ensures robustness against both gross rotation errors and subtle scanning distortions. Correcting skew is essential to prevent systematic erros from OCR recognition and the subsequent summary or extraction tasks.  

\textbf{Re-normalization:} Finally, pages are standardized to a fixed resolution and aspect ratio to ensure consistency across downstream OCR processing. We rescale cropped pages using Bicubic interpolation \cite{han2013comparison}, which preserves sharpness in small characters and table borders. To further enhance readability, we apply Contrast Limited Adaptive Histogram Equalization (CLAHE) \cite{reza2004realization} for local contrast normalization and light Gaussian denoising to suppress background artifacts such as stains or shadows. This combination improves text–background separation and ensures numerical entries remain crisp for reliable downstream OCR recognition.  

Overall, this pre-processing pipeline substantially improves OCR robustness, especially for heterogeneous SMB reports that vary in layout quality, image sizes, and scanning conditions.

\subsection{Multi-lingual OCR Transcription}
After pre-processing, each page is passed through an OCR system to recover textual content. We employ PaddleOCR v3 \cite{cui2025paddleocr}, a state-of-the-art lightweight OCR framework, due to its strong multilingual coverage, speed, and robustness for noisy scanned documents.

In particular, PaddleOCR incorporates two main modules. (i) \textbf{Text detection:} A differentiable bounding-box detector locates text regions at multiple scales, which is particularly important for financial reports containing mixed layouts of narrative paragraphs, tables, and marginal notes. (ii) \textbf{Text recognition:} The detected regions are transcribed into character sequences using language-specific recognition heads. PaddleOCR supports over 80 languages, including English, Simplified Chinese, Traditional Chinese, etc., in both printed and handwritten format. The multilingual features, together with printed/handwritting variance supports, are essential in processing SMB financial disclosures.  

\begin{figure}
  \centering
  \includegraphics[width=\linewidth]{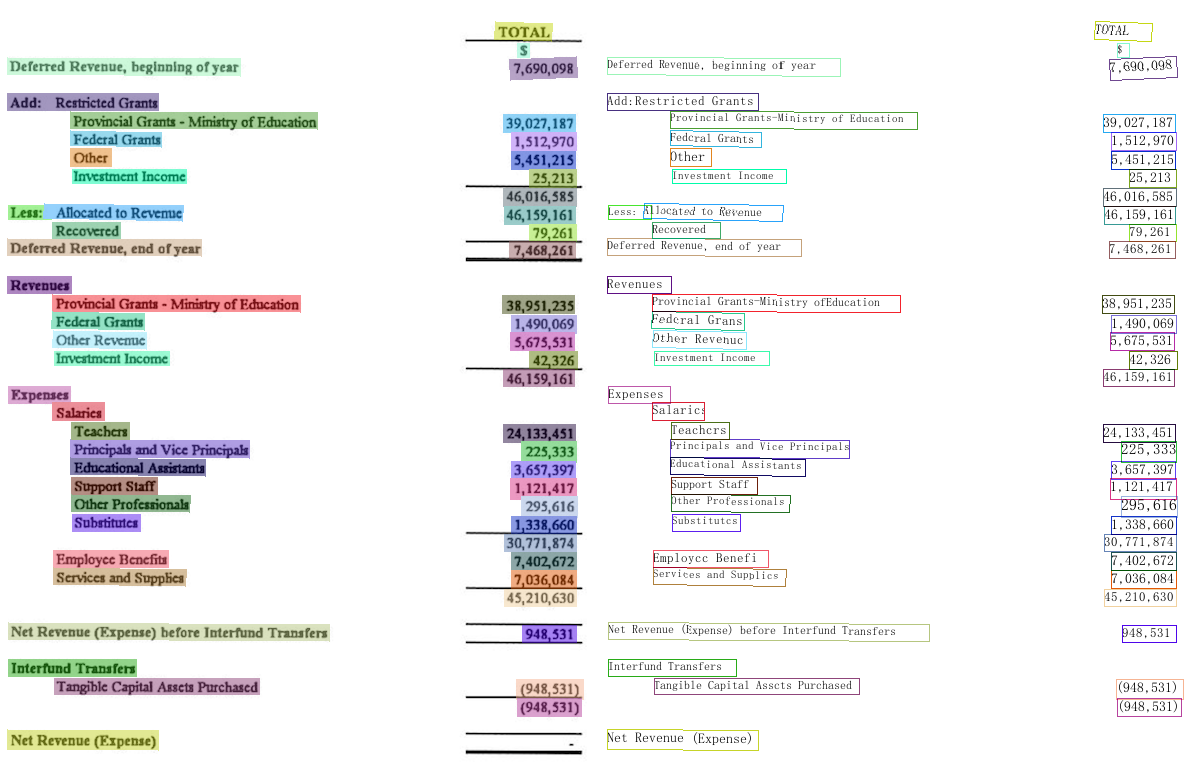}
  \caption{Sample OCR transcription output showing recognized text, confidence scores, and bounding box coordinates for layout-aware extraction.}
  \Description{OCR transcription}
  \label{fig:ocr}
\end{figure}

Figure \ref{fig:ocr} demonstrates the sample output of the OCR transcription. Specifically, for each recognized token, the OCR system outputs not only the transcribed text but also a confidence score along with spatial coordinates of the bounding box. These metadata allow downstream components to filter out low-confidence entries and preserve layout-sensitive structures such as tables. 

By leveraging PaddleOCR v3, we ensure reliable text transcription in heterogeneous multilingual reports, providing a solid foundation for the downstream tasks.

\subsection{Page Retrieval}
Following OCR transcription, the next step is to identify the subset of pages that are most relevant for extracting target financial fields. Processing the entire document with a Vision-Language Model (VLM) would be computationally prohibitive, especially for reports spanning hundreds of pages. Instead, we introduce a lightweight page retrieval stage to narrow down the search space.  

We employ a BM25-based keyword retrieval method \cite{robertson2009probabilistic}, a classical information retrieval model that ranks documents (in our case, pages) according to their lexical similarity to a query. For each pre-defined target field (e.g., revenue, profit, dividends, executive compensation), we define a small set of representative keywords and phrases. Pages are indexed using their OCR-transcribed tokens, and BM25 is used to assign a relevance score to each page based on keyword frequency and contextual weighting.  

This approach has four main advantages. (i) \textbf{Efficiency:} By filtering out irrelevant sections such as company background or general risk disclosures, we significantly reduce the number of pages passed into downstream VLMs. (ii) \textbf{Robustness:} BM25’s ranking function tolerates OCR noise, since high-frequency keywords can still dominate even when transcription errors occur. (iii) \textbf{Flexibility:} New target fields can be added simply by expanding the set of keywords, without retraining or modifying the retrieval component. (iv) \textbf{Lightweight:} BM25 requires minimal computational resources compared to LLM-based or embedding-based retrieval, and executes with much lower latency. This makes the approach well suited for GPU-constrained in-house deployments.  

Note that we also experimented with popular text embedding-based Retrieval Augmented Generation (RAG) methods, where dense vector embeddings are computed for each page and compared against query vectors. However, financial reports present a particular challenge that the textual summaries of individual pages often contain very similar content (e.g., repeated mentions of accounting terms and disclaimers). This leads embedding-based methods to produce nearly uniform similarity scores across pages. In contrast, BM25 leverages explicit keyword frequency, making it more effective in isolating specific financial indicators such as net profit versus other types of profit, total revenue versus all revenue sources, or dividend per share versus other dividends. 

\begin{figure*} [t]
  \centering
  \includegraphics[width=0.9\linewidth]{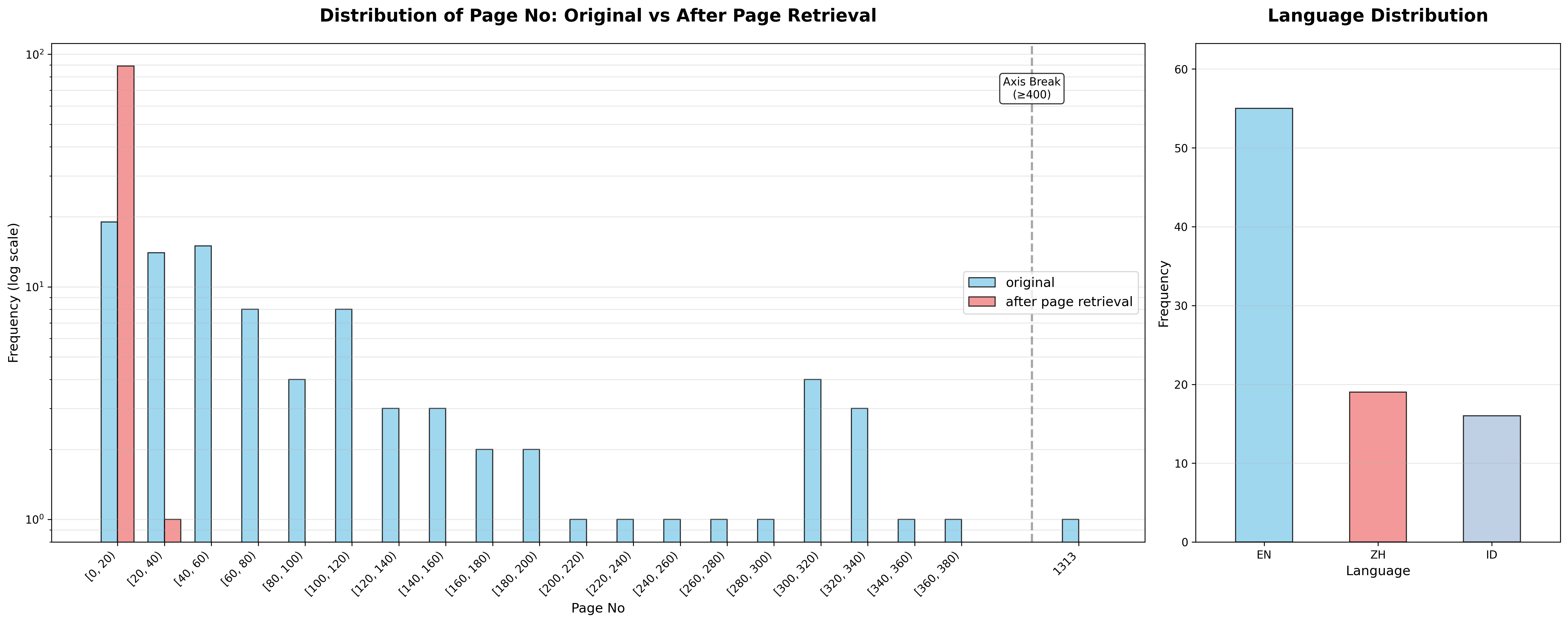}
  \caption{Characteristics of the in-house experimental dataset with 93 scanned financial documents from SMBs. (i) Left: document lengths distribution (before and after page retrieval). (ii) Right: document language distribution.}
  \Description{page distribution}
  \label{fig:dataset}
\end{figure*}

\subsection{Extraction using Compact VLM}
The final stage of our pipeline applies compact VLMs to extract structured financial fields from the retrieved pages. The dramatically narrowed down sections identified in the retrieval stage, allows smaller VLMs to operate effectively within their limited context windows, while still capturing the multimodal relationships between text and layout from the targeted pages.  

During this step, compact VLMs are prompted to extract predefined fields of interest, such as revenue, net profit, dividends, and executive compensation. Since the inputs include both the OCR text and images with highlights on the textual layouts, the models are capable of handling complex structures such as tables or mixed text-table regions. In addition, the use of section-specific prompts improves reliability by tailoring the extraction process to the expected content type and its language.  

This approach offers three benefits. (i) GPU memory and inference time are significantly reduced compared to large VLMs, enabling practical deployment in resource-constrained in-house environments. (ii) It mitigates noise sensitivity, since the models operate on cleaner and pre-filtered inputs rather than noisy full-document streams. (iii) the modular extraction design improves interpretability. Each extracted value can be traced back to its origin page and region, providing the possibility of straightforward human-in-the-loop verifications.

Our experiments evaluated two settings, including miniCPM-o 2.6 \cite{yao2024minicpm} as the compact VLM (one 80GB Nvidia A100 card can host six miniCPM replicas together with 6 PaddleOCR replicas) and Qwen2.5-VL-72B-Instruct \cite{bai2025qwen2} as the large VLM (at least two 80GB  Nvidia A100 cards can host only one 72B QWen replica).

\section{Experiments}
\label{section:experiments}
In this section, we present an empirical evaluation of the proposed multi-stage parsing framework. We first describe the in-house dataset of financial documents (including company financial statements and 3rd-party auditing reports) used for benchmarking. Next, we outline the evaluation methodology, including the target fields, baseline models, and performance metrics. We then report the results of our experiments, comparing compact VLMs with large VLM baselines in terms of accuracy, efficiency, and latency. Finally, we conduct an error analysis to identify the main sources of failure and discuss their implications for future improvements.

\subsection{Dataset}
The experimental dataset consists of \textbf{92} scanned financial reports or 3rd-party audit reports collected from SMBs, covering multiple languages and document formats from different countries. As shown in Figure \ref{fig:dataset}-(i), the original reports vary significantly in length, ranging from fewer than 20 pages to over 1,000 pages. After applying page retrieval, most documents are reduced to under 20 relevant pages, yielding an average reduction ratio of approximately 92\% in the number of processed pages. This substantially lowers the computational cost of downstream components, allowing more focused extraction with significantly improved performance. Figure \ref{fig:dataset}-(ii) illustrates the language distribution, which is dominated by English (56 reports), followed by Chinese (19 reports) and Indonesian Bahasa (17 reports). This diversified data set allows us to assess both the scalability of the pipeline across long documents and the robustness of the OCR, page retrieval, and information extraction components under multipage and multilingual conditions.

Note that this internal data set is highly confidential and contains real customer information that we could not disclose. The figures we illustrate in this paper are all publicly accessible, and look similar in terms of scan quality, key structure, and length of the documents. But they are not the real ones from our data set.     

\subsection{Evaluation Approach}
To systematically assess the effectiveness of our proposed pipeline, we designed an evaluation framework that measures both \textbf{accuracy of information extraction} and \textbf{system efficiency}. Specifically, we consider the following dimensions.

\begin{figure}
  \centering
  \includegraphics[width=\linewidth]{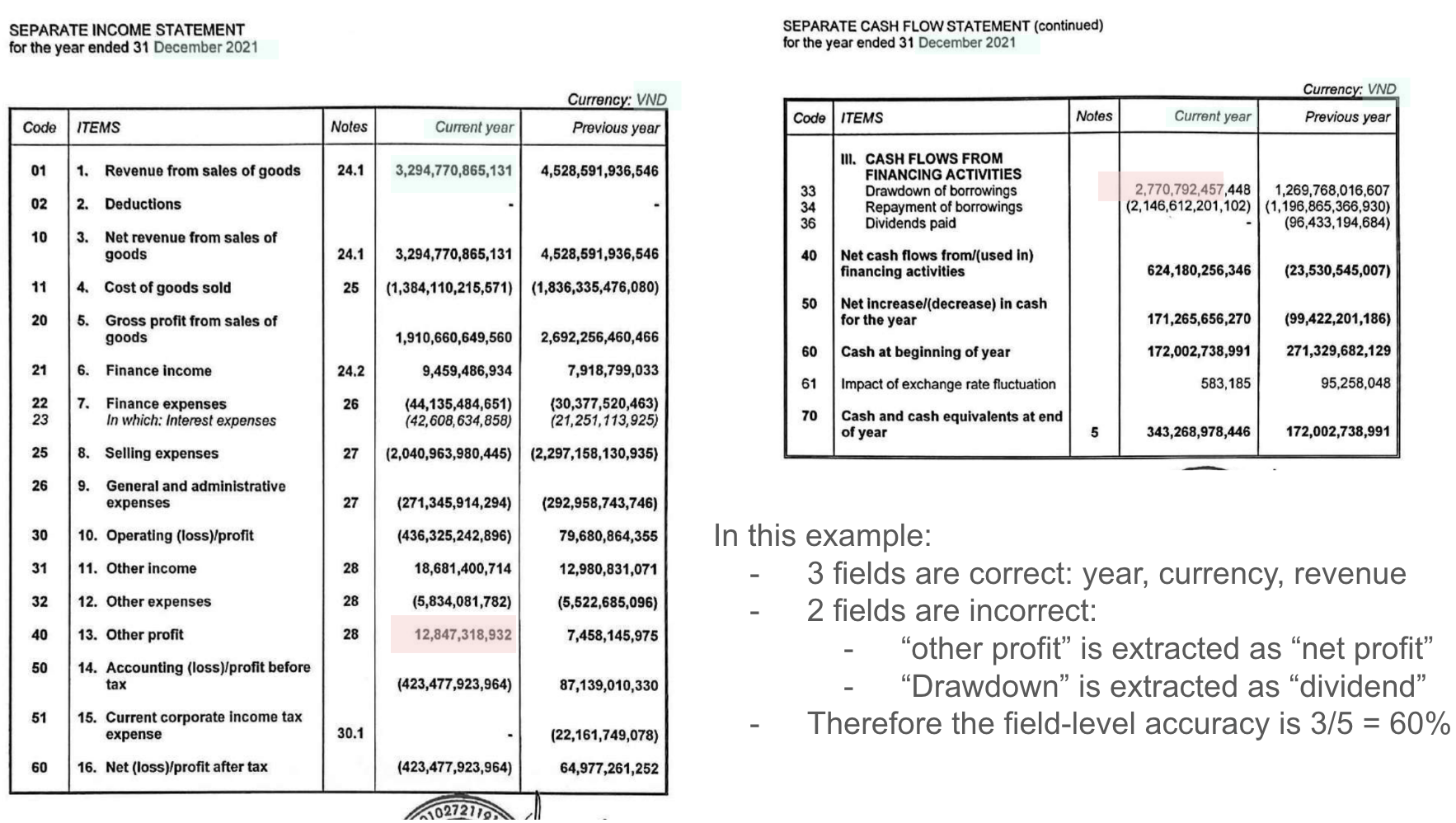}
  \caption{Field-level accuracy calculation example based on five pre-defined fields.}
  \Description{Accuracy example}
  \label{fig:example}
\end{figure}

\begin{itemize}
    \item \textbf{Extraction Accuracy:} We evaluate field-level accuracy, indicating whether key financial fields (in particular five fields including year, revenue, profit, dividends, and currency) are correctly extracted compared to manually curated ground truth. Specifically, figure \ref{fig:example} illustrates an example of how we calculate such field-level accuracy in this study.  
    
    \item \textbf{Model Efficiency:} To quantify computational cost, we measure GPU throughput in terms of the number of documents that can be processed per hour per Nvidia A100 GPU card.
    
    \item \textbf{Latency in Service Context:} For in-house deployment scenarios, we measure response latency from document input to structured output. This metric reflects user-facing efficiency and is critical when GPU resources are limited. To make a fair comparison between short documents with only dozens of pages and very long documents with hundreds of pages, we report the average latency per page.  
\end{itemize}

This evaluation not only demonstrates the accuracy of structured field extraction but also highlights the computational savings and practical feasibility of our approach in real-world financial services.

\subsection{Results}
Table~\ref{tab:results} presents the end-to-end performance comparison between a one-shot extraction baseline using QWen2.5-VL-72B as the large VLM, and our proposed multi-stage pipeline leveraging OCR and several pre-processing steps, together with miniCPM-o-2.6-8B as the compact VLMs. The results highlight several key findings.  

First, the one-shot large VLM baseline achieved only 9\% field-level accuracy. Most of such tries for documents with too many pages, ended up with Out-Of-Memory (OOM) errors, without any valid extractions to be returned. Even when restricted to short documents with fewer than 15 pages, the model was still often distracted by a large proportion of irrelevant content (e.g., company background or boilerplate risk disclosures). Without page retrieval, this irrelevant information overwhelms the model’s attention, leading to significant degradation in  extraction accuracy.  

Second, our multi-stage pipeline with compact VLMs dramatically improved performance, reaching around 80\% field-level accuracy. This represents a \textbf{8.8 times} accuracy boost, demonstrating the effectiveness of combining keyword-driven retrieval and localized extraction to ensure the VLM only processes semantically relevant pages. By narrowing the input scope, the pipeline reduces error propagation and makes efficient use of VLM model capacity. 

Third, the efficiency benefits are substantial at only \textbf{0.7\%} of the GPU cost. The proposed pipeline achieved a throughput of 1,522 documents per hour per A100, compared to only 11 documents per hour per A100 for the one-shot large VLM. This significant gap comes from multiple factors, including (i) the lightweight OCR steps to narrow down the scopes for more time-consuming VLM calls. (ii) the considerable size differences between large and compact VLMs. Specifically, when considering GPU resource allocation, a single A100 card can host up to six MiniCPM (8B) replicas as well as six PaddleOCR replicas, enabling parallelized batch inference. In contrast, a single Qwen model requires at least two A100 cards to host just one replica, severely limiting deployment scalability.  

Finally, latency per page was reduced from 9.656 seconds to 0.717 seconds, representing \textbf{92.6\%} latency reduction, even after excluding the error cases. This improvement can be mostly attributed to the much faster OCR processing time to quickly filter out irrelevant pages, comparing to the overwhelming inference time by blindly applying large VLM into the whole document. In addition, the parallelized OCR and compact VLM processing capabilities under the same GPU resources further widen the gap. 

Taken together, these results highlight that compact VLMs, when embedded within a retrieval and pre-processing pipeline, not only achieve significantly higher accuracy than large VLMs, but also enable cost-effective and scalable deployment in real-world financial workflows with limited in-house GPU resources.

\begin{table}[t]
\centering
\caption{End-to-end performance comparison between one-shot extraction using large VLM and the proposed multi-stage pipeline using compact VLM. The later one achieves significant improvements in terms of both accuracy and GPU resource consumption and latency.}
\label{tab:results}
\begin{threeparttable}
\begin{tabular}{|p{2.1cm}|c|p{1.9cm}|c|}
\hline
\textbf{} & \textbf{Accuracy $\uparrow$} & \textbf{Throughput $\uparrow$} & \textbf{Latency $\downarrow$}  \\ \hline
One-shot extraction using large VLM & 9.13 \% & 11 docs/h per A100 & 9.656 s/page \tnote{1} \\ \hline
Multi-stages pipeline using compact VLM & \textbf{80.87 \%}  & \textbf{ 1,522 docs/h per A100} & \textbf{0.717 s/page} \\ \hline
\end{tabular}
\begin{tablenotes}
\footnotesize
\item[1] The latency for one-shot large VLM call already excludes the OOM error cases, and based on only the successful cases from <10\% cases.
\end{tablenotes}
\end{threeparttable}
\end{table}

\subsection{Error Analysis}
Despite the strong performance of our proposed multi-stage pipeline, several sources of error remain, contributing to incorrect extracted fields. We summarize the key observations below.  

\subsubsection{Inconsistent Terminology.}  
A large source of error arises from inconsistent naming conventions across different audit firms. For example, the same financial concept may be referred to as \emph{revenue} in one report and \emph{income} or \emph{sales} in another. Similarly, \emph{dividend} may be ambiguously interchanged with \emph{interest}, especially when dealing with multi-lingual sources. While different types of profit, such as profit before tax, profit excluding tax, net profit, etc., would easily confuse the compact VLM doing extraction. These variations cause keyword-based retrieval and compact VLM extraction to occasionally mismatch or overlook the intended fields.

\subsubsection{Currency Unit Ambiguities.}  
Currency units, particularly in Indonesian financial statements, present another significant challenge. For instance, amounts may be reported as \emph{"IDR'000"} or \emph{"IDR'000,000"}, expressed in natural text as \emph{"ribuan rupiah"} (thousand rupiah), or as \emph{"juta rupiah"} (million rupiah). Such inconsistencies complicate normalization, as the same numeric figure can represent values differing by three to six orders of magnitude depending on the context. Although compact VLMs are able to capture part of this context, automated disambiguation remains non-trivial and may require additional learning-based normalization.

\subsubsection{OCR Errors.}  
Another chunk of errors are contributed from OCR. In particular, the OCR step may fail to capture critical keywords, primarily due to low-quality scans or overlapping stamps obscuring the text. While the BM25 retrieval module is robust to minor transcription noise, complete keyword omission can prevent the correct page from being retrieved, leading to missed extractions.

Overall, these errors highlight the need for improved robustness in both the retrieval and extraction stages, particularly through semantic normalization of terminology and currency units, and integration of domain-specific lexicons to reduce variability across reporting styles among different languages and countries.

\section{Conclusion and Future Works}
\label{section:conclusion}
In this study, we have shown that with a multi-stage pipeline combining traditional image pre-processing, OCR transcription, page retrieval, and compact VLM extraction, it is feasible to accurately extract structured financial data from large, noisy, multilingual scanned documents. The approach offers strong efficiency and interpretability, making it more efficient and practical for real-world finance applications in large scale.

Our future work aims to construct a much larger in-house dataset with thousands of more diversified financial documents for more comprehensive evaluations. On top of that dataset, more carefully designed ablation studies will be conducted to better understand not only the specific impact of each individual step along the pipeline but also to reach more confident conclusions.

\begin{acks}
This work was conducted at OCBC AI Lab, together with colleagues from various departments across OCBC and its private banking arm, Bank of Singapore (BOS). The authors would like to specifically thank Kelvin Chiang, Ngee Hai Heng, Srinivasan Thangamani, Rionaldi Chandraseta, Addison Chew, Kent Moi, Kelvin Heng, and Wentao Xu for their valuable contributions and supports throughout the project. Their insights and feedback were instrumental in shaping the research and experiments presented in this paper.
\end{acks}

\bibliographystyle{ACM-Reference-Format}
\bibliography{reference}










\end{document}